\documentclass[twocolumn, prb,superscriptaddress,amsfonts,showpacs,floatfix]{revtex4} 
\usepackage{color}

\usepackage{graphicx}
\usepackage{dcolumn}
\usepackage{bm}

\begin{document}

\preprint{APS/123-QED}

\title{
Unconventional Strong Spin-Fluctuation Effects around the Critical Pressure of 
the Itinerant  Ising-Type Ferromagnet URhAl  
}

\author{Yusei Shimizu }
\email{yuseishimizu@issp.u-tokyo.ac.jp}
\affiliation{Univ. Grenoble Alpes, INAC-SPSMS, CEA-Grenoble, F-38000 Grenoble, France. }
\author{ Daniel Braithwaite }
\affiliation{Univ. Grenoble Alpes, INAC-SPSMS, CEA-Grenoble, F-38000 Grenoble, France. }
\author{Bernard Salce}
\affiliation{Univ. Grenoble Alpes, INAC-SPSMS, CEA-Grenoble, F-38000 Grenoble, France. }
\author{Tristan Combier}
\affiliation{Univ. Grenoble Alpes, INAC-SPSMS, CEA-Grenoble, F-38000 Grenoble, France. }
\author{ Dai Aoki}
\affiliation{Univ. Grenoble Alpes, INAC-SPSMS, CEA-Grenoble, F-38000 Grenoble, France. }
\author{ Eduardo N. Hering}
\affiliation{Univ. Grenoble Alpes, INAC-SPSMS, CEA-Grenoble, F-38000 Grenoble, France. }
\author{ Scheilla M. Ramos}
\affiliation{Univ. Grenoble Alpes, INAC-SPSMS, CEA-Grenoble, F-38000 Grenoble, France. }
\author{Jacques Flouquet}
\affiliation{Univ. Grenoble Alpes, INAC-SPSMS, CEA-Grenoble, F-38000 Grenoble, France. }


\date{\today}

\begin{abstract}
Resistivity measurements   were performed for the itinerant Ising-type ferromagnet URhAl
  at  temperatures down to 40 mK under high pressure up to 7.5 GPa, 
 using  single crystals.
We found that the  critical pressure of the Curie temperature exists at around $P_{  \mathrm{c} } \sim$ 5.2 GPa.
Near  $P_{ \mathrm{c} }$, 
 the $A$-coefficient of the $AT^{2}$ Fermi-liquid resistivity term below $T^{*}$ is largely enhanced 
 with a maximum around 5.2-5.5 GPa.
\color{black}
Above  $P_{\mathrm{c} }$,  
 the exponent of  the  resistivity $\rho(T)$  deviates from 2.
 At $P_{\mathrm{c}}$, it is
 close to $n = 5/3$, which is  expected by the theory of   three-dimensional ferromagnetic spin fluctuations for a 2nd-order quantum-critical point (QCP).  
However, $T_{\mathrm{C} }(P) $ disappears as a 1st-order phase transition, and 
 the critical behavior of resistivity in URhAl cannot be explained  by the theory of a 2nd-order QCP.  
The 1st-order nature of the phase transition is weak, and 
 the critical behavior is still dominated by the spin fluctuation at low temperature.
With increasing pressure, 
 the  non-Fermi-liquid behavior is observed in higher  fields.
 Magnetic field studies point out a ferromagnetic wing structure with a  
 tri-critical point (TCP) at $\sim$ 4.8-4.9 GPa in URhAl.
One open possibility
 is that the switch from the ferromagnetic to the paramagnetic states does not  occur simply but 
 an intermediate state arises below the TCP as suggested theoretically recently.  
Quite generally, if a drastic Fermi-surface 
 change occurs through $P_{\mathrm{c} }$, the nature of the 
 interaction itself may change and lead to the observed unconventional behavior.
\color{black}
\end{abstract}

\pacs{71.27.+a, 75.30.Kz, 75.30.Mb, 75.40.-s}
\maketitle

\section{Introduction}

Since the 1960-1970s, the understanding of  dynamic 
  critical phenomena  and  physical properties  
  of itinerant  spin-fluctuation  systems 
  has   been  one of the main topics in the fields of  magnetism 
  in condensed matter physics    
   \cite{Doniach_PRL_1966, Berk_PRL_1966, Beal-Monod_PRL_1969,
  Moriya_PRL_1970, Moriya_JPSJ_1973, Moriya_JPSJ_1995}.
This is  because these questions lead to understand  
 not only weak  itinerant magnetism  in $d$- and $f$-electron systems  
  but also recently observed anomalous non-Fermi-liquid  (NFL) 
 \color{black} behaviors 
  and  magnetically mediated Cooper instabilities  
 \cite{SpinFluctuation_Moriya, Moriya_RepProgPhys_2003}
  caused  by spin fluctuations near  quantum-phase transitions (QPTs).

So far, it was widely believed that both itinerant ferromagnetic (FM) and antiferromagnetic (AF) compounds usually have the quantum-critical points (QCPs), 
 where a 2nd-order phase transition occurs at $T = 0$ by tuning some physical parameter, such as pressure, or atomic substitution, etc. 
The self-consistent-renormalized (SCR) theory by Moriya and his coworkers gives a theoretical base to describe NFL behaviors of itinerant FM and AF metallic systems near QCPs 
\cite{  Moriya_JPSJ_1995, SpinFluctuation_Moriya, Moriya_RepProgPhys_2003}.
 Furthermore, critical phenomena around magnetic QCPs were investigated theoretically using the renormalization-group method by Hertz 
\cite{Hertz_PRB_1976}, and Millis  \cite{Millis_PRB_1993}.
 Actually, some itinerant AF compounds obey the Moriya-Hertz-Millis theory for critical behaviors near QCPs \cite{Lohneysen_RevModPhys_2007}.

\color{black}

However, for FM quantum criticality, the situation is different.
Surprisingly, 
 it has been  reported that an almost FM helimagnet, MnSi
 \cite{Pfleiderer_PRB_1997, Pfleiderer_Nature_2001}, and several 
 ferromagnets, such as UGe$_{2}$
 \cite{Pfleiderer_PRL_2002,
 Taufour_PRL_2010}, and ZrZn$_{2}$
 \cite{Uhlarz_PRL_2004, Kabeya_JPSJ_2012}, do not show the QCP at zero field but show a 1st-order phase transition  
  when $T_{\mathrm{C} }$ is suppressed by applying pressure.
To explain these behaviors,
 recently  specific attentions were given to new quantum treatment for FM QPT; 
 for example, 
 \color{black}
 Belitz and Kirkpatrick considered particle-hole excitations from a Fermi surface with a low frequency and a long wavelength (soft mode),
 which couples to the order parameter
 \cite{Belitz_PRL_1999, Belitz_PRL_2005}.
They  showed that 
  a 2nd-order phase transition at high temperature changes to a 1st-order transition  
 below a \textit{tri-critical point} (TCP) with     
1st-order \textit{wing} planes, which terminate 
at zero temperature in a finite magnetic field, i.e. at a quantum-critical-end point (QCEP)
  \cite{Belitz_PRL_1999, Belitz_PRL_2005}.
Previously, 
 it has also been discussed that the TCP emerges due  solely to  
 the thermal spin fluctuations  
 \cite{Yamada_PRB_1993, Yamada_PhysicaB_2007}
 and the magneto elastic coupling 
 \cite{Mineev_JPhysConfSer_2012, Gehring_EurophysLett_2008}. 

So far,  the quantum criticality around the QCEP with the metamagnetic 
 transition  has been classified into the same criticality as the QCP for an Ising-type  transition
 \cite{Millis_PRL_2002}.
However, 
 there is no symmetry change around a QCEP, whereas 
   the symmetry of the ordered phase is clearly distinguished from 
 the paramagnetic (PM) phase for a QCP.
It has recently been 
 pointed out theoretically that the quantum criticality of the  metamagnetic transition accompanied with the Fermi-surface change (Lifshitz transition) has another universality class, which differs from  other symmetry-breaking phase transitions
 \cite{Yamaji_JPSJ_2007,  Imada_JPhysC_2010}.
Also, 
 as unconventional superconductivity associated with FM order has been discovered only in uranium materials (UGe$_{2}$
 \cite{Saxana_Nature_2000}, 
 URhGe \cite{Aoki_Nature_2001}, 
 and UCoGe \cite{Huy_PRL_2007}), 
  it is  intriguing to study 
  the quantum criticality and the spin-fluctuation effects around the FM QPT for itinerant uranium compounds.

\color{black}

\begin{centering}
\begin{figure}[!htb]
\includegraphics[width=7.7cm]{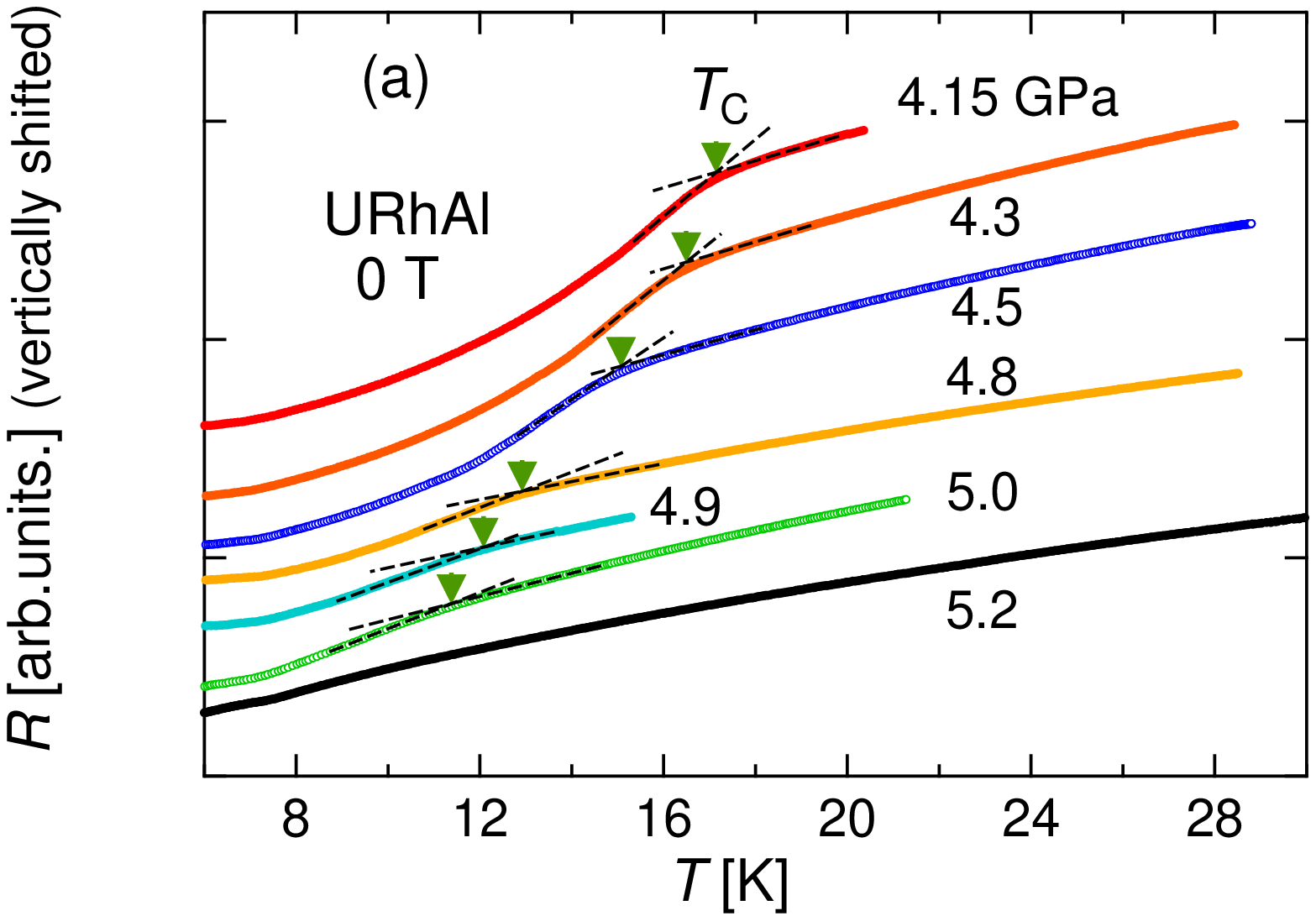}
\includegraphics[width=7.7cm]{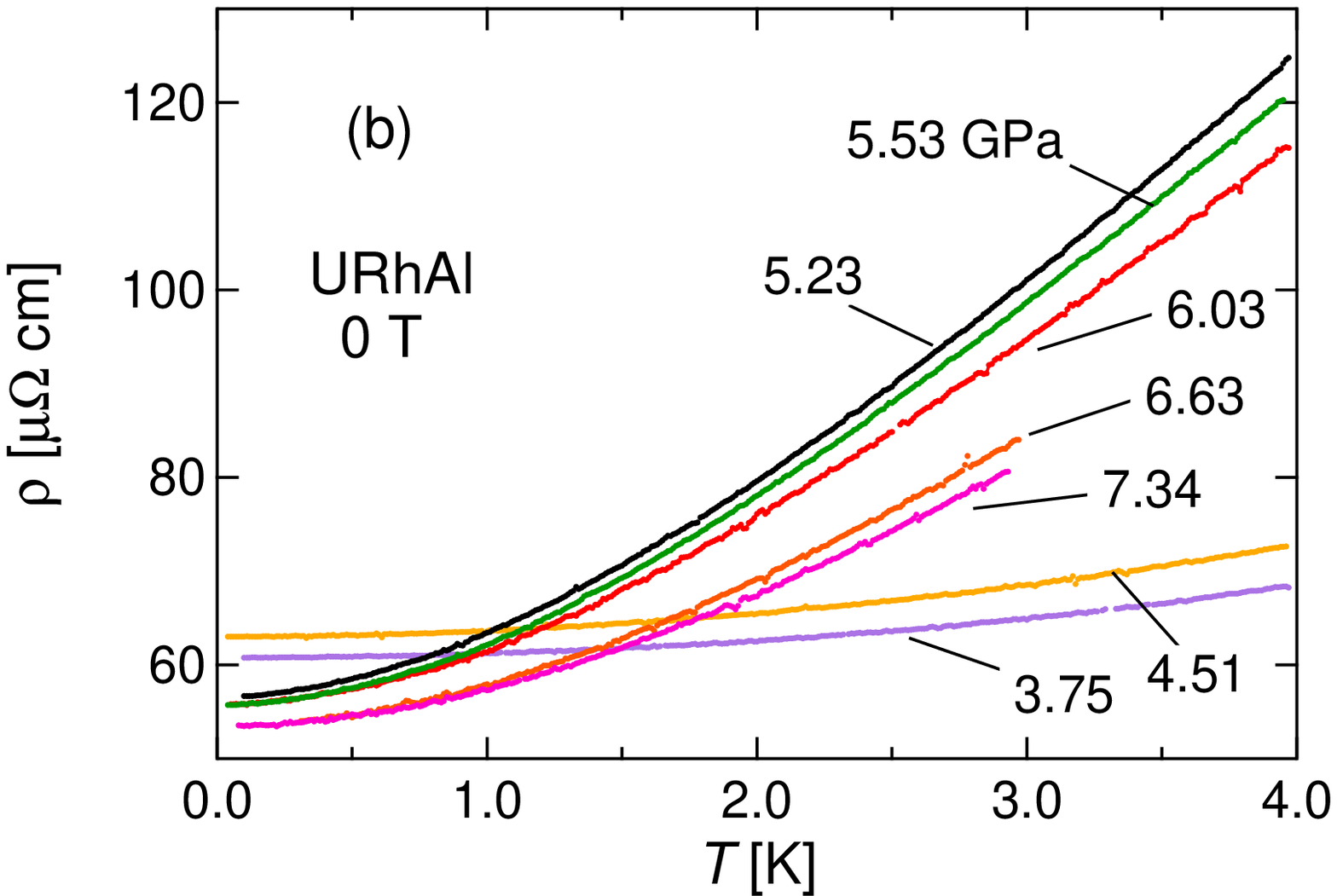}
\caption{ (Color online)
(a) Temperature dependence of resistance (vertically shifted) of  URhAl (sample $\#1$) between 6 and 30 K, measured at zero field and high pressures, 4.15, 4.3, 4.5, 4.8, 4.9, 5.0, and 5.2 GPa.
The  arrows indicate the Curie temperatures ($T_{\mathrm{C}}$)  at each pressure. The dashed lines are guides to the eyes.
(b) Temperature dependence of resistivity  of  URhAl (sample $\#1$) below 4 K,  
measured at zero field and high pressures, 3.75, 4.51, 5.23, 55.3, 6.03, 6.63, and 7.34 GPa.
}
\end{figure}
\end{centering}

  Recently, a FM wing structure and a QCEP
 have been reported for UCoAl, 
 which shows   a 1st-order metamagnetic transition
 at  $\sim$ 0.7 T with a FM moment of $\sim$ 0.3 $\mu_{\mathrm{B} }/$U at low temperature 
 \cite{Andreev_SovPhys_1985, Matsuda_JPSJ_1999, Aoki_JPSJ_2011}.
This  compound has  a hexagonal ZrNiAl-type structure with space group $P\bar{6}2m$,
 in which there is no inversion symmetry.
The uranium atoms form a quasi-Kagom\'{e} lattice, thus 
 magnetic frustration effects are possibly expected.
From high-pressure studies, 
it is considered that in UCoAl a TCP exists at negative pressure of $-$0.2 GPa
 \cite{Mushnikov_PRB_1999}, and 
 the  metamegnetic transition can be explained  by the  FM wings 
 \cite{Aoki_JPSJ_2011}.
Since the TCP in UCoAl is estimated to exist at a negative pressure,  
 it is not observable from hydrostatic-pressure measurements. 
In order to understand the critical phenomena near the itinerant FM QPTs, 
further experimental examples are necessary.

In this paper,   
 we report pressure-induced quantum criticality of a $5f$ itinerant 
  Ising-type FM compound, URhAl, which has  the same crystal structure as that of  UCoAl.
URhAl shows a FM transition at 25-27 K at ambient pressure
 \cite{Veenhuizen_J_de_Physique_1988, Javorsky_PRB_2004, TristanCombier_Dr_Thesis_2014}, and 
  the FM moment ($\sim$ 0.9 $\mu_{\mathrm{B} }$/U) is strongly  Ising-type with the magnetization-easy axis along  $c$,   similar to the Ising-type metamagnetism in UCoAl.
The atomic radius of Rh is larger than that of Co, so  the  
 $5f$-electronic state in URhAl  may correspond  to a state in which  negative pressure is applied for UCoAl
 \cite{Andreev_JAlloysComp_1999}.
Therefore, the high-pressure study of critical behaviors for URhAl  can help 
   to understand the metamagnetism in UCoAl as well as the general problem of   FM quantum criticality.   

\section{Experimental Procedures}

Single crystals of URhAl  were grown by the Czochralski pulling method in a tetra-arc furnace.
Resistivity measurements  under high pressure were performed by using diamond-anvil cells 
  with an  \textit{in situ} pressure-tuning device 
 \cite{Salce_RevSciInstrum_2000, Demuer_JLTP_2000}.
We measured resistivity of samples 
 ($\#1$ and $\#2$) which were less than the size of $\sim$ 200 $\times$ 100 $\times$ 30 $\mu$m$^{3}$.
The sample geometry did not allow a precise 
 determination of the form factor of resistivity.
Therefore, 
 we extrapolated $A(P)$ linearly to 0 GPa, and 
 obtained absolute values of $\rho(T,H)$ and $A$ 
 by  normalizing the extrapolated value [$A(P = 0)$]  to 
 the zero-pressure value ($A$ = 0.27 $\mu \Omega $ cm/K$^{2}$ for $J \perp c$),
 since  the pressure variation of $A$-coefficient is almost linear for $P < $4.8 GPa.

The low-$T$  measurements were carried out for sample $\#1$  
 using a $^{3}$He-$^{4}$He dilution refrigerator 
in temperatures down to  40 mK and in fields up to 7 T under high pressures up to 7.5 GPa.
Here, the magnetic field was applied almost along the 
 $c$-axis (easy-magnetization axis) and the current was applied perpendicular to the field direction.
The high-$T$  measurements under high pressure were  performed  at zero magnetic  field using $^{4}$He cryostat 
 for sample $\#1$ as well as $\#2$ to check the reproducibility. 

As a pressure transmitting medium,
  liquid argon was used,
 and pressure was determined with the ruby fluorescence technique.
For high-$T$ measurements,  since there is a volume increase of helium gas in bellows of the pressure-tuning 
 device  above liquid-helium temperature,  
  this may cause a slight change of the force, which is applied to the pressure cell.
Then, 
 the determination of pressure is more precise for low-$T$ measurements below $\sim$ 4 K than for high-$T$ measurements 
 above $\sim$ 5 K.

\section{Results and Discussion}

 In order to examine the pressure dependence of the Curie temperature of URhAl,
 we first show  the temperature dependence of the resistance at various pressures 
 (shifted vertically)
 between 6 and 30 K in Fig. 1(a).
One can see the clear kink anomaly in the resistivity curves due to 
 the FM transition at the Curie temperature  ($T_{\mathrm{C} }$),
 as indicated by the  arrows [Fig. 1(a)].
 $T_{\mathrm{C} }$ shifts
 to lower temperature with increasing pressure,
 and  the kink anomaly  becomes 
 too broad to determine $T_{\mathrm{C} }$ for $P >$ 5.0 GPa.

Figure 1(b) shows results of resistivity measurements below  4 K
 at high pressures from 3.75 to 7.34 GPa.
At  3.75 and 4.51 GPa, $T_{\mathrm{C}}$ is   19 and  17 K, respectively,
 and   URhAl is FM  in the temperature range of Fig. 1(b) at these pressures. 
The  variation of resistivity $\rho(T)$ is small at low temperature in the FM state.
On the other  hand,
 from  5.2 to 7.3 GPa,
 the variation of resistivity is 
 very large compared to that at  3.75 and 4.51 GPa.
Since we did not observe the kink anomaly  in the resistivity due to the FM transition above  5.2 GPa,
 URhAl   appears to be PM in  the high-pressure region above  5.2 GPa.

\begin{centering}
\begin{figure}[!htb]
\includegraphics[width=8.6cm]{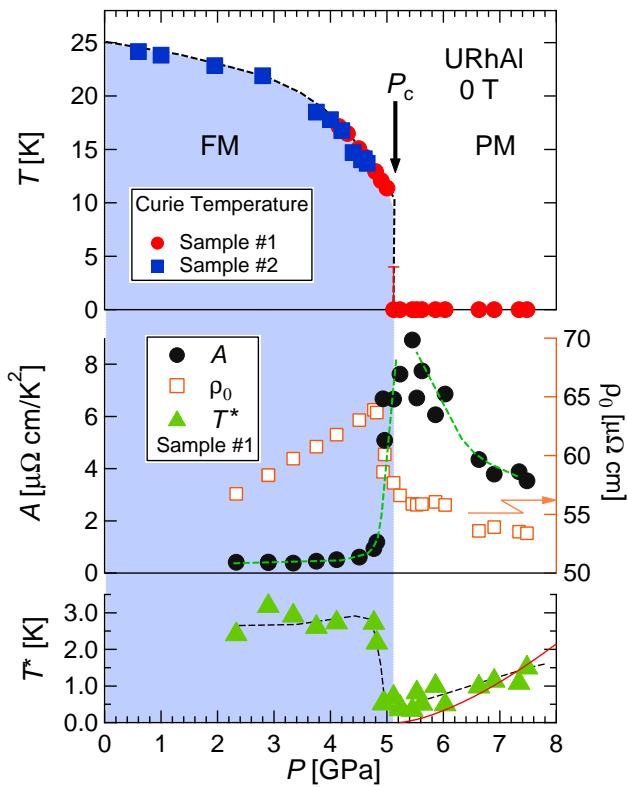}
\caption{ (Color online) $T$-$P$ phase diagram of URhAl at zero field for the  samples $\#1$ and  $\#2$. 
The dashed lines are guides to the eyes.
The $A$-coefficient and the residual resistivity of the sample $\#1$ are also plotted.
The solid  curve indicates $T^{*}(P) \propto (P -P_{\mathrm{c} })^{3/2}$, which is predicted by the spin-fluctuation theory for a 2nd-order FM QCP
 \cite{Millis_PRB_1993, Flouquet_ArXiv_2005}.  
\color{black}
}
\end{figure}
\end{centering}

Figure 2 shows the obtained  $T$-$P$ phase diagram at zero field.
 There is no significant sample dependence of  $T_{\mathrm{C} }(P)$.
 $T_{\mathrm{C} }(P)$ suddenly 
disappears above  5.0 GPa.
Our results suggest that 
the FM critical pressure exists at $P_{\mathrm{c}} \sim $ 5.2 GPa.

In Fig. 2, we also plot the $A$-coefficient and the residual resistivity $\rho_{0}$
 as a function of pressure, where $\rho(T) = \rho_{0} + AT^2$.
 The pressure variation of 
 the 
 \color{black}
$A$-coefficient is very small ($A \sim$ 0.4-0.5 $\mu \Omega $cm/K$^{2}$) for $P < $ 4.8 GPa,
 whereas it shows a drastic increase above  5.0 GPa.
The $A$-coefficient becomes a maximum ($A \sim$ 9 $\mu \Omega$cm/K$^{2}$) at around 5.2-5.5 GPa,
 suggesting a large  enhancement of the density of states at Fermi energy [$D(\epsilon_{\mathrm{F} })$] of 
 itinerant  electrons above $P_{\mathrm{c} }$.
%
%
Interestingly, the large increase 
 of the $A$-coefficient ($A \sim $4 $\mu \Omega $cm/K$^{2}$) beyond $P_{\mathrm{c} }$
 indicates that the large enhancement of $D(\epsilon_{\mathrm{F} })$ 
 remains up to $\sim$ 7.5 GPa (Fig. 2).
The behavior of the residual resistivity, $\rho_{0}(P)$, 
 accompanies the behavior of the $A$-coefficient, $A(P)$.
Below 4.8 GPa, $\rho_{0}(P)$ increases with increasing pressure almost linearly,
  then  suddenly decreases with increasing pressure above  4.9 GPa.
At 0 T
 $\rho_{0}(P)$ shows a step-like behavior at $\sim$ 4.8 GPa  slightly below $P_{\mathrm{c} } \sim $ 5.2 GPa.

We also plot $T^{*}$,  the maximum temperature for the $T^{2}$-regime,  in the 3rd panel of  Fig. 2.
 $T^{*}$ is about $\sim$ 2-3 K in the low-pressure region below $P_{\mathrm{c} }$,
 but it suddenly decreases at $P_{\mathrm{c}}$, where 
  $T^{*}(P)$ shows a minimum ($T^{*} \sim $ 0.4 K), and then gradually 
 increases with increasing pressure.

 In FM spin fluctuation frame with a 2nd-order QCP,
   $T^{*}(P)$ and $A(P)$ are predicted to vary as $T^{*}(P) \propto (P - P_{\mathrm{c} })^{3/2}$ and $A(P) \propto (P - P_{\mathrm{c} })^{-1}$, 
respectively, leading to $A \propto (1/T^{*})^{2/3}$;
 in other words, 
 $A \times (T^{*})^{2/3}$ is constant
 \cite{Millis_PRB_1993, Flouquet_ArXiv_2005}. 
For URhAl, 
 we obtain $A \sim 8$ $\mu \Omega$cm/K$^{2}$ and $T^{*} \sim 0.4 $ K at  $P_{\mathrm{c} }$, leading to $A \times (T^{*})^{2/3} \sim$ 4.3, and 
 $A \sim 3.5$ $\mu \Omega$cm/K$^{2}$ and $T^{*} \sim 1.5 $ K at  7.5 GPa, leading to $A \times (T^{*})^{2/3} \sim$ 4.6.
This rough  estimation  
 suggests that the observed large $A$-coefficient  emerges due to the FM spin-fluctuation effects. 
However,
 we would like to point out the peculiarity of critical behavior of the FM QPT in URhAl;
   as shown in the 3rd panel of Fig. 2,    
 $T^{*}(P) $ does not vary as    $T^{*}(P) \propto (P - P_{\mathrm{c} })^{3/2}$  (the solid curve) and 
 does not  go to zero as $P \rightarrow P_{\mathrm{c} }$.
Also, the fact that $T^{*}$ is finite at $P_{\mathrm{c} }$ conflicts with  presence of a 2nd-order QCP in URhAl.

\color{black}

\begin{table}
\caption{The $A$-coefficient of resistivity, $A$ [$\mu \Omega$cm/K$^2$], the electronic specific-heat 
 coefficient $\gamma$ [mJ/K$^{2}$mol], and the values of $A/\gamma^{2}$ [$\mu \Omega$ cm$ (\mathrm{mol K})^2/(\mathrm{mJ} )^2$] 
 for URhAl,  UCoAl\cite{Aoki_JPSJ_2011}, and UGe$_{2}$\cite{Tateiwa_JPhysC_2001}.
 Here, $A(0)$ is the value of $A$-coefficient at ambient pressure at zero field.
}
\begin{ruledtabular}
\def\arraystretch{1.3}
\begin{tabular}{lcrrrr}
                                 & $P$ [GPa]                        & $A$     & $\gamma$ &  $A/\gamma^{2}$     & $A/A(0)$ \\
\hline
URhAl                         &  0         (FM)                     &  0.27   & 75            & 4.8$\times$10$^{-5}$ &   $-$   \\
                                &  $P_{\mathrm{c} } \sim $5.2   &   8      &               &                              &    $30$ \\
\hline
UCoAl                        &  0                                     &  0.28   &  75           & 5.0$\times$10$^{-5}$ &  $-$    \\
                                &  0.54                                 &  0.2     &              &                        &  0.7     \\
                                & $P_{\mathrm{QCEP}}\sim$1.5, 7 T &  0.4    &             &                        &  1.4     \\
\hline
UGe$_{2}$                   &  0                                     &  0.007  &  30           &  7.8$\times$10$^{-6}$ &   $-$ \\
                                 &  1.3                                 &    0.1   &    110       &  8.3$\times$10$^{-6}$ &   14.3   \\
\end{tabular}
\end{ruledtabular}
\end{table}

\begin{centering}
\begin{figure}[!htb]
\begin{minipage}{4.2cm}
\includegraphics[width=4.2cm]{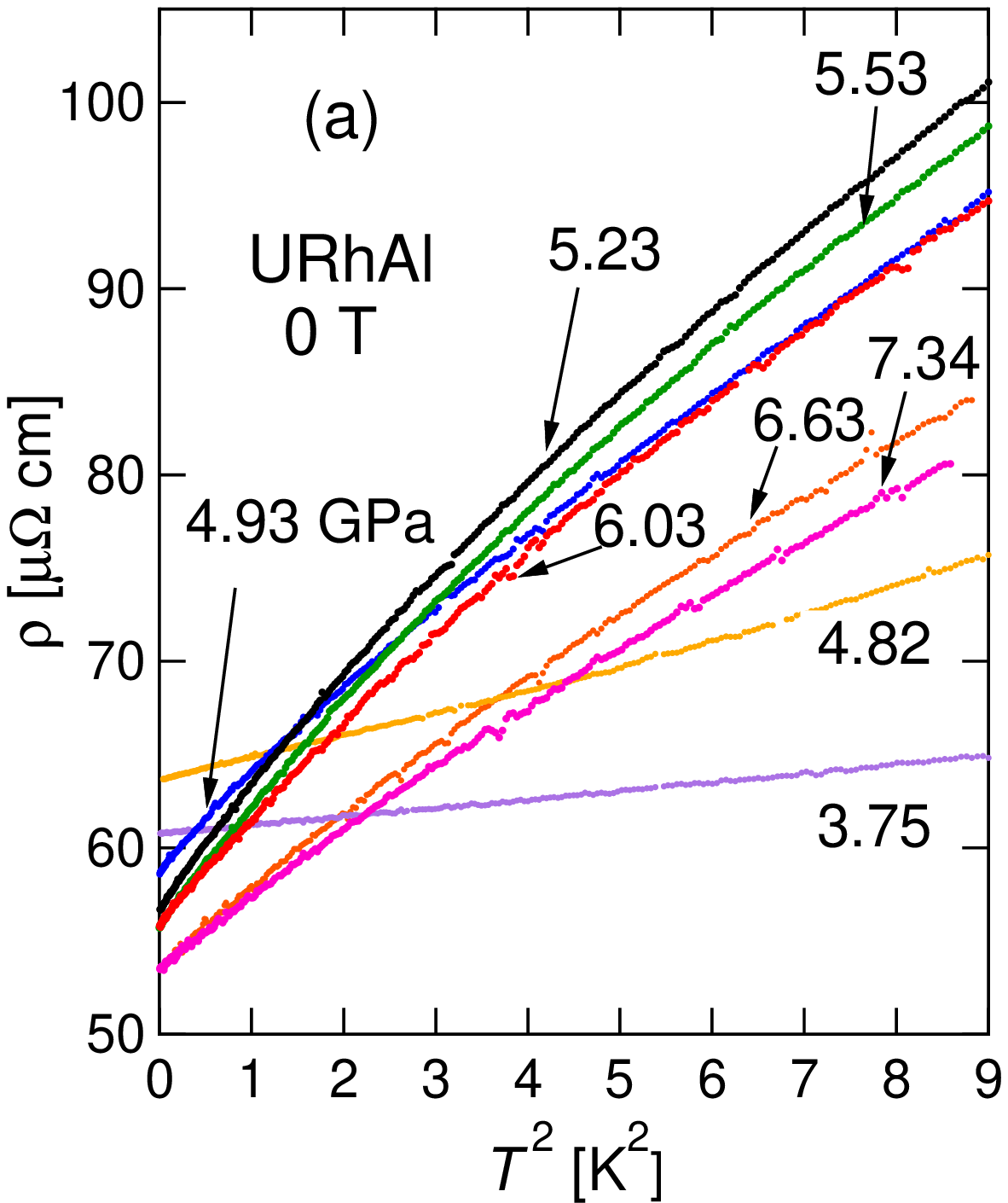}
\end{minipage}\hspace{0cm}%
\begin{minipage}{4.2cm}
\includegraphics[width=4.2cm]{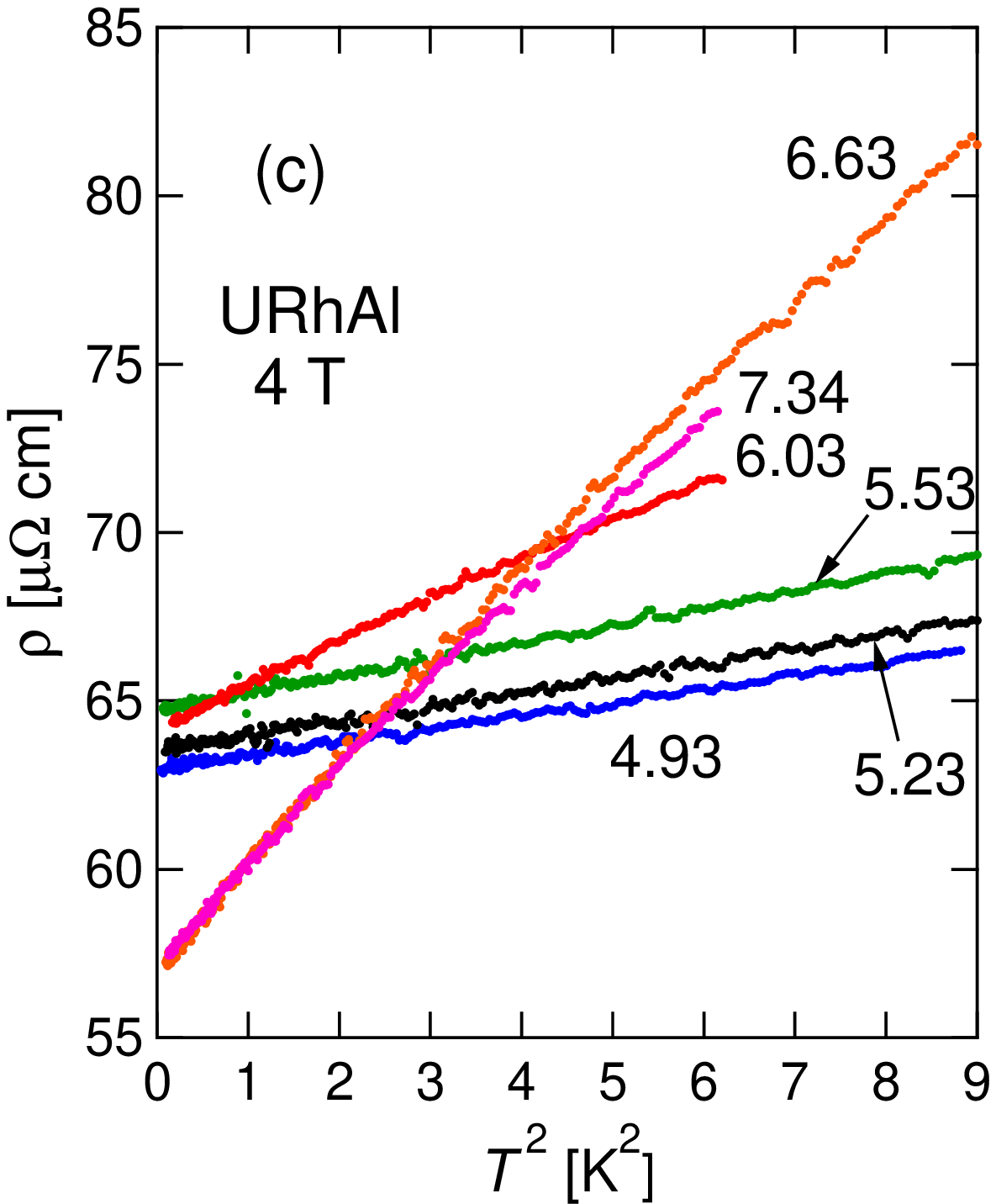}
\end{minipage} 
\begin{minipage}{4.2cm}
\includegraphics[width=4.2cm]{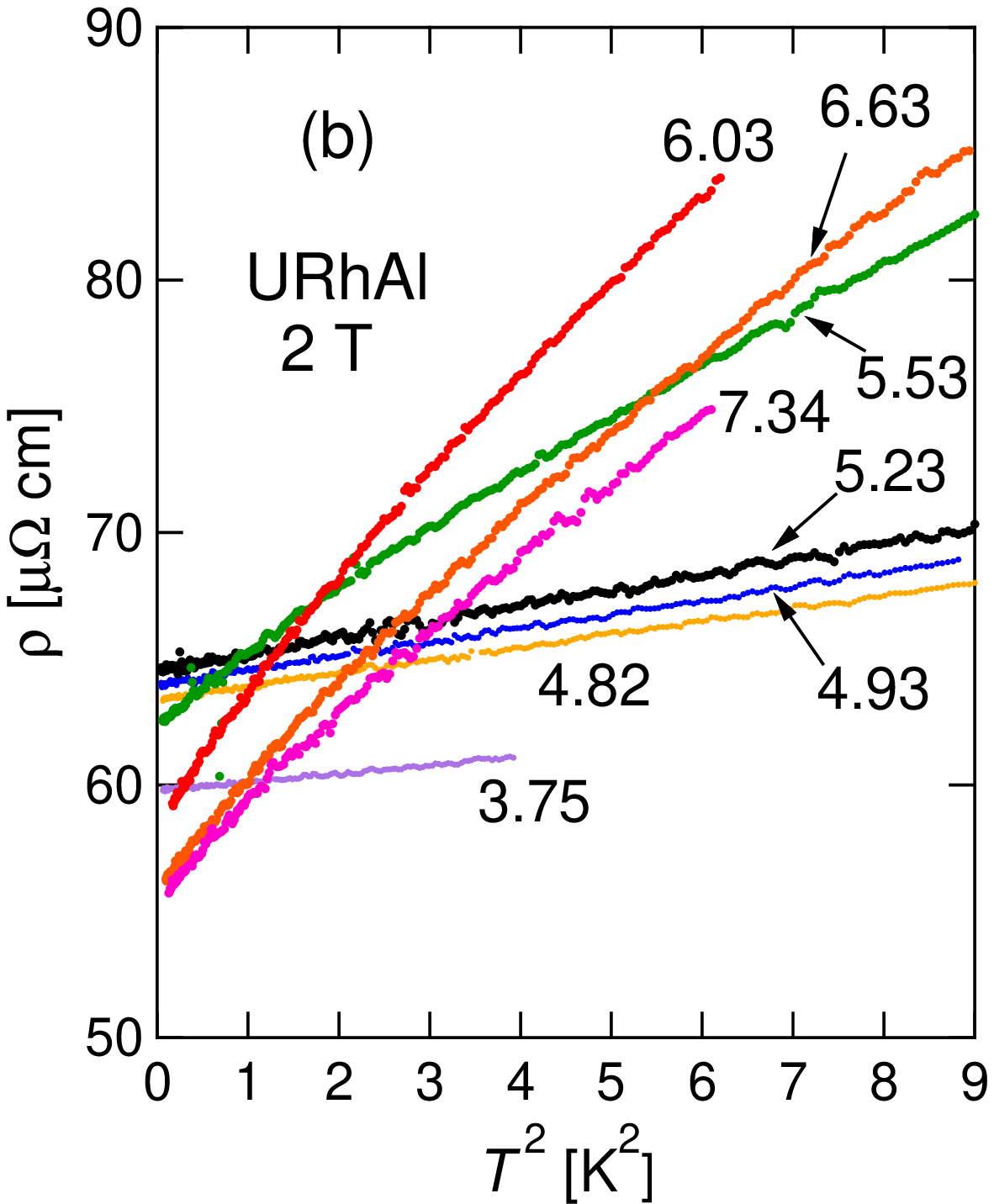}
\end{minipage}\hspace{0cm}%
\begin{minipage}{4.2cm}
\includegraphics[width=4.2cm]{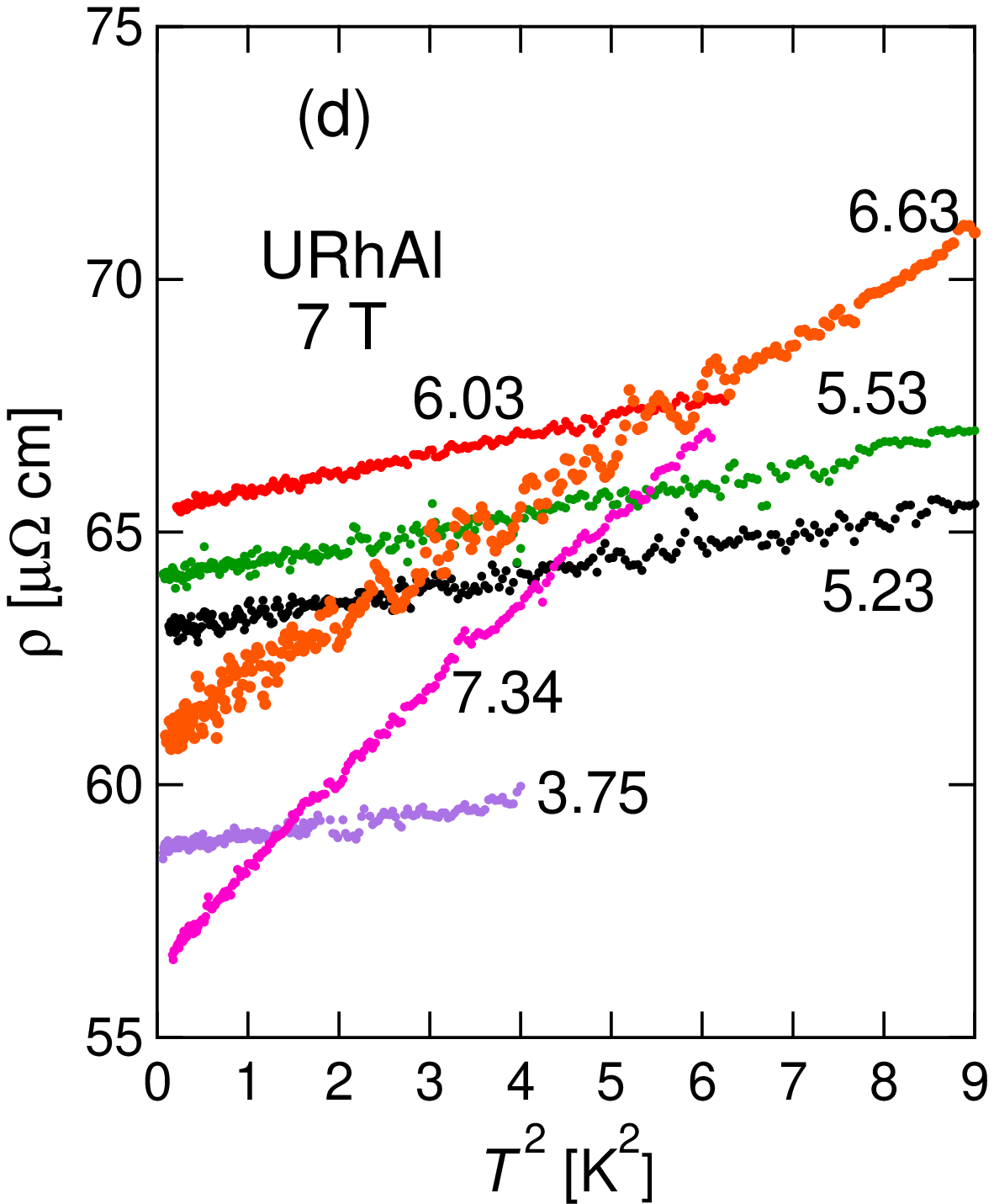}
\end{minipage} 
\includegraphics[width=8cm]{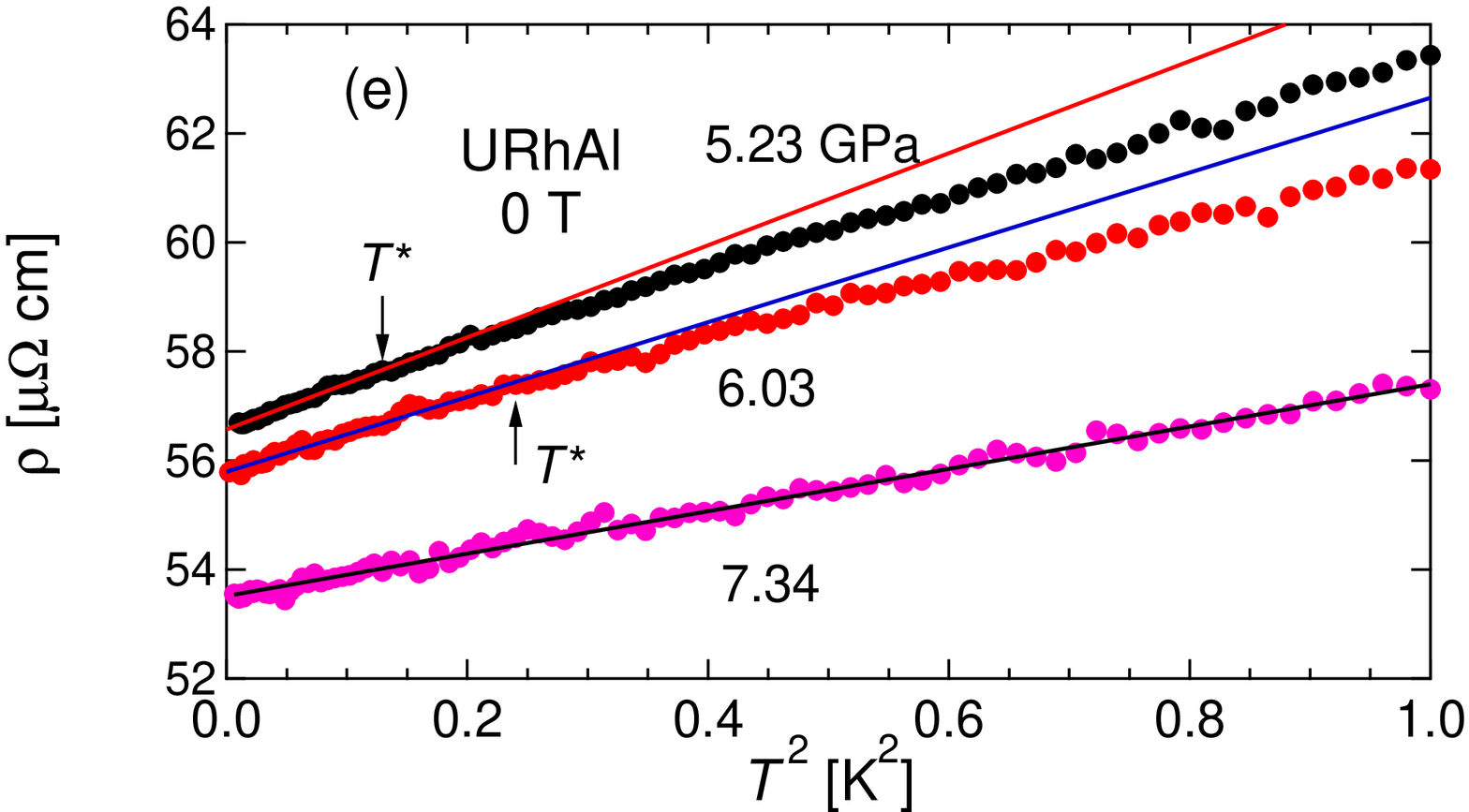}
\caption{ (Color online) $\rho(T)$ vs $T^{2}$ plot of URhAl (sample $\#1$) at high pressures from 3.75 to 7.34 GPa in
  (a) 0 T, (b) 2 T, (c) 4 T, and (d) 7 T.
(e) The enlarged figure of $\rho(T)$ curves  for
0 T measured at 5.23, 6.03, and 7.34 GPa as a function of $T^2$ below 1 K.
The arrows indicate $T^{*}$ (Fig.2), below which
the $T^2$-regime works.
 Here, $T^{*}$ at 7.34 GPa is about $\sim$ 1.1 K.
The solid lines are the results of fitting by $\rho(T) = \rho_{0} + AT^2$.
\color{black}
}
\end{figure}
\end{centering}

\color{red}
\color{black}
The  maximum value of $A$ $\sim$ 9 $\mu \Omega$cm/K$^{2}$ in URhAl near $P_{\mathrm{c}}$   
 is quite large for uranium intermetallic compounds.
While the heavy-electron superconductor UBe$_{13}$ shows the 
exceptionally large $A$-coefficient ($\sim$ 90-100 $\mu \Omega$cm/K$^{2}$)
 \cite{Remenyi_JPhysique_1986, KodowakiWoods_SolidStateCom_58_1986},
 a lot of uranium compounds
 show the $A$-coefficient of less than  $\sim$ 1 $\mu \Omega$ cm/K$^{2}$
 as summarized in the Kadowaki-Woods plot 
 \cite{KodowakiWoods_SolidStateCom_58_1986}.
Table I shows the $A$-coefficient, the electronic specific-heat coefficient ($\gamma$), and 
 the ratio  $A/\gamma^{2}$ for URhAl \cite{TristanCombier_Dr_Thesis_2014},
  UCoAl \cite{Aoki_JPSJ_2011}, and UGe$_{2}$ \cite{Tateiwa_JPhysC_2001}. 
Besides,  $A/A(0)$ indicates the ratio of $A$ divided by the $A$-coefficient at 0 GPa and zero field, i.e. $A(0)$. 
 As for UCoAl, the $A$-coefficient is 
 $A \sim 0.28$ $\mu \Omega$ cm/K$^{2}$, and the electronic specific-heat coefficient is $\gamma \sim$ 75
 mJ/K$^2$mol
 \cite{Aoki_JPSJ_2011}.
The $A$-coefficient of UCoAl increases near the QCEP ($\sim$ 1.5 GPa, 7 T)
 \cite{Aoki_JPSJ_2011},
 but the enhancement  of $A$ is not so large compared to 
 the pressure-induced large $A$-coefficient in URhAl.
Also, the $A$-coefficient of  UGe$_{2}$ increases $\sim$14-fold under high pressure,
 but the maximum value of  $A$-coefficient is not so large ($\sim$ 0.1 $\mu \Omega$ cm/K$^{2}$)
 at $\sim$1.3 GPa 
 \cite{Tateiwa_JPhysC_2001}.
On the other hand, 
 a large $A$-coefficient ($\sim$ 5 $\mu \Omega$cm/K$^{2}$) near the critical pressure
 has been reported in  an itinerant heavy-electron FM compound U$_{4}$Ru$_{7}$Ge$_{6}$ 
 \cite{Hidaka_JPSJ_2011}.
The observed large $A$-coefficient in URhAl near $P_{\mathrm{c} }$
 is comparable with the value observed in cerium heavy-electron compounds 
 such as CeCu$_{2}$Si$_{2}$ 
\cite{KodowakiWoods_SolidStateCom_58_1986, Holmes_PRB_2004}.
From the comparison with other heavy-electron materials using the Kadowaki-Woods relation,
 the quantum critical region in URhAl 
 may be described grosso modo by  the strongly correlated heavy quasiparticle
 with the large $D(\epsilon_{\mathrm{F}})$ caused by spin fluctuations.
However, 
 we should be careful about the 
 above discussion since the value of $A/\gamma^2$ is not universal 
 depending on the correlation of the system
 \cite{EndoNote_KadowakiWoods, Miyake_SolidStateCom_1989, Morales_Thesis_2014}.
\color{black}

\begin{centering}
\begin{figure}[!htb]
\includegraphics[width=7.8cm]{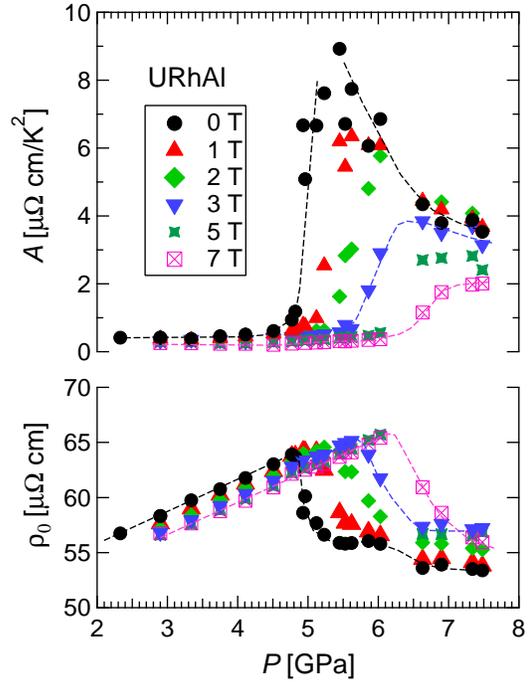}
\caption{(Color online) Pressure dependence of the $A$-coefficient and the residual resistivity $\rho_{0}$ of URhAl (sample $\#1$) in zero and magnetic fields, obtained from 
 the  expression, $\rho(T) = \rho_0 + AT^2$. 
The dashed lines are guides to the eyes.
}
\end{figure}
\end{centering}

Next, we shall see low-$T$ $\rho(T)$ curves in zero field and magnetic fields.
Figures 3(a), (b), (c), and (d) show
 the resistivity   $\rho(T)$ vs $T^{2}$   under high pressures from
 3.75 to 7.34 GPa
 for 0, 2, 4, and 7 T,  respectively.
For zero field, 
  at lower pressures than 4.8 GPa, 
 we find   $\rho(T) = \rho_{0} + AT^{2}$ behavior, as predicted for  an 
 itinerant FM state at low temperature ($T$ $\ll$ $T_{\mathrm{C} }$)
 \cite{Ueda_JPSJ_1975}.
On the other hand,   
 the resistivity shows a remarkable 
 variation with 
  a large increase of the slope ($A$-coefficient) at 0 T 
 between 4.82 and 4.93 GPa.
Around 5-6 GPa, 
 the temperature region where the resistivity can obey the expression 
 $\rho(T) = \rho_{0} + A T^2$ is  much smaller than at   
 4.82, and 3.75 GPa.
 In Fig. 3(e), we show the enlarged figure of $\rho(T)$ curves for 0 T 
measured at 5.23, 6.03, and 7.34 GPa as a function of $T^2$. The arrows indicate the temperature, $T^{*}$ (Fig.2), below which the $T^2$-regime works.
\color{black}
For an applied field of 2 T, the large $A$-coefficient  
 is suppressed  at 4.93, and 5.23 GPa, and $\rho(T)$ shows the  $T^{2}$-temperature dependence, similar to that at 4.82, and 3.75 GPa.
On the other hand, the slope of $\rho(T)$ becomes large at around 6-6.6 GPa at 2 T.
For 4 T, and 7 T,  
 the variation of $\rho(T)$ at high-pressure region above $\sim$ 6.6 GPa becomes larger than for pressures below $\sim$ 6.0 GPa.

In Fig. 4, we summarize the  pressure dependence of the $A$-coefficient and the residual resistivity $\rho_{0}$ of URhAl (sample $\#1$), obtained from 
 the  expression, $\rho(T) = \rho_0 + AT^2$,
 in zero and magnetic fields. 
With increasing magnetic field, 
 the divergence of the $A$-coefficient is suppressed, and the step-like behavior
 of $\rho_{0}$ becomes broad  (Fig. 4).

Since the behaviors of the $A$-coefficient and $\rho_{0}$ above   5.0 GPa
  differ evidently from those below $\sim$ 5.0 GPa in the FM state,
 it is considered that URhAl   
 is  \textit{not} in the FM state any more  above 5.0 GPa.
This  is consistent with the fact 
that the anomaly due to the FM transition disappears above  5.0 GPa 
 (Fig. 2).
The Curie temperature, $T_{\mathrm{ C}}(P)$, possibly becomes a 1st-order phase transition,   
 and then  $T_{\mathrm{C} }(P)$ suddenly collapses above  5.0 GPa.

\begin{figure}[!htb]
\includegraphics[width=7.8cm]{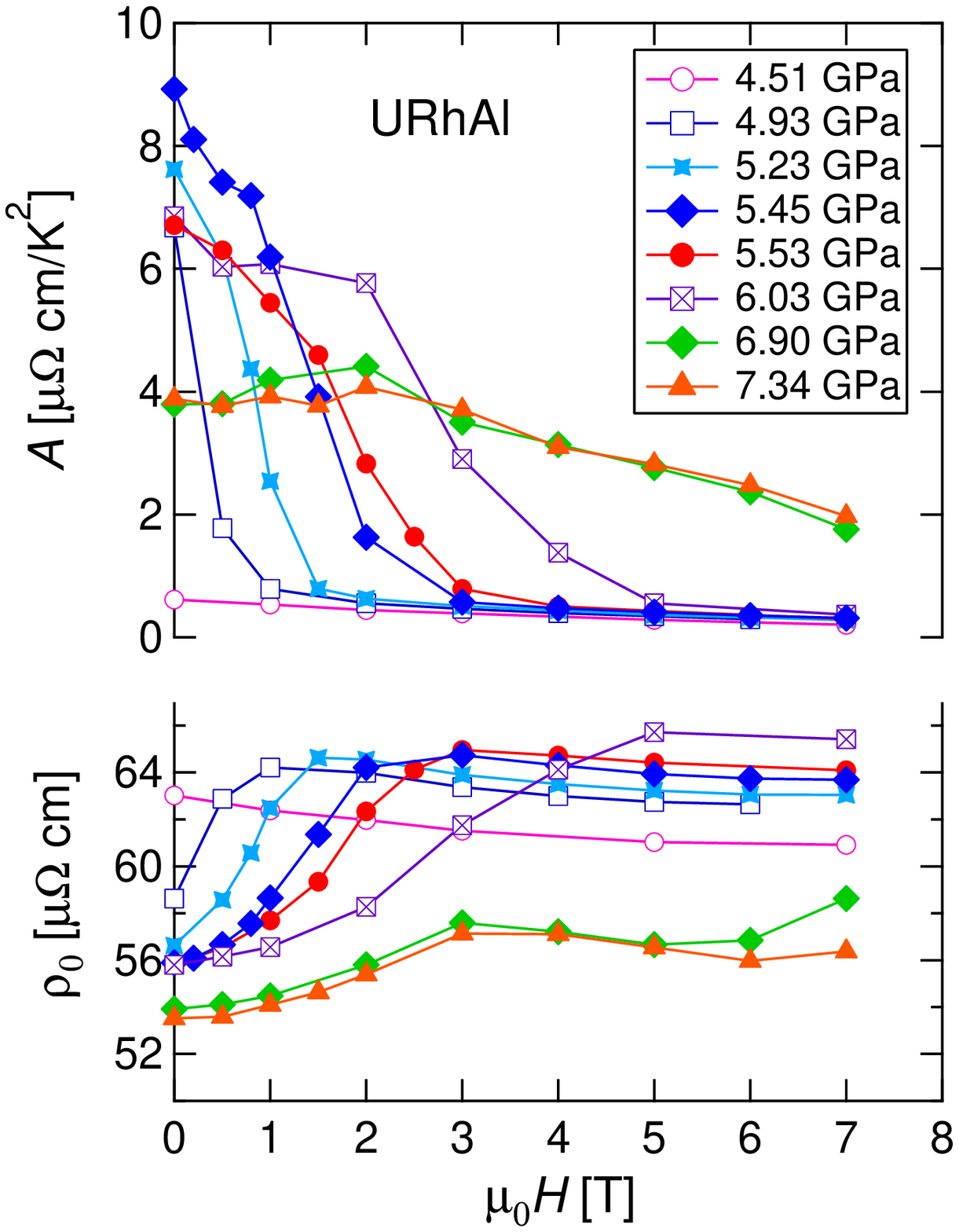}
\caption{ (Color online) Magnetic-field dependence of the $A$-coefficient and the residual resistivity $\rho_{0}$ of URhAl (sample $\#1$), obtained from 
 the  expression, $\rho(T) = \rho_0 + AT^2$.  
}
\includegraphics[width=7.5cm]{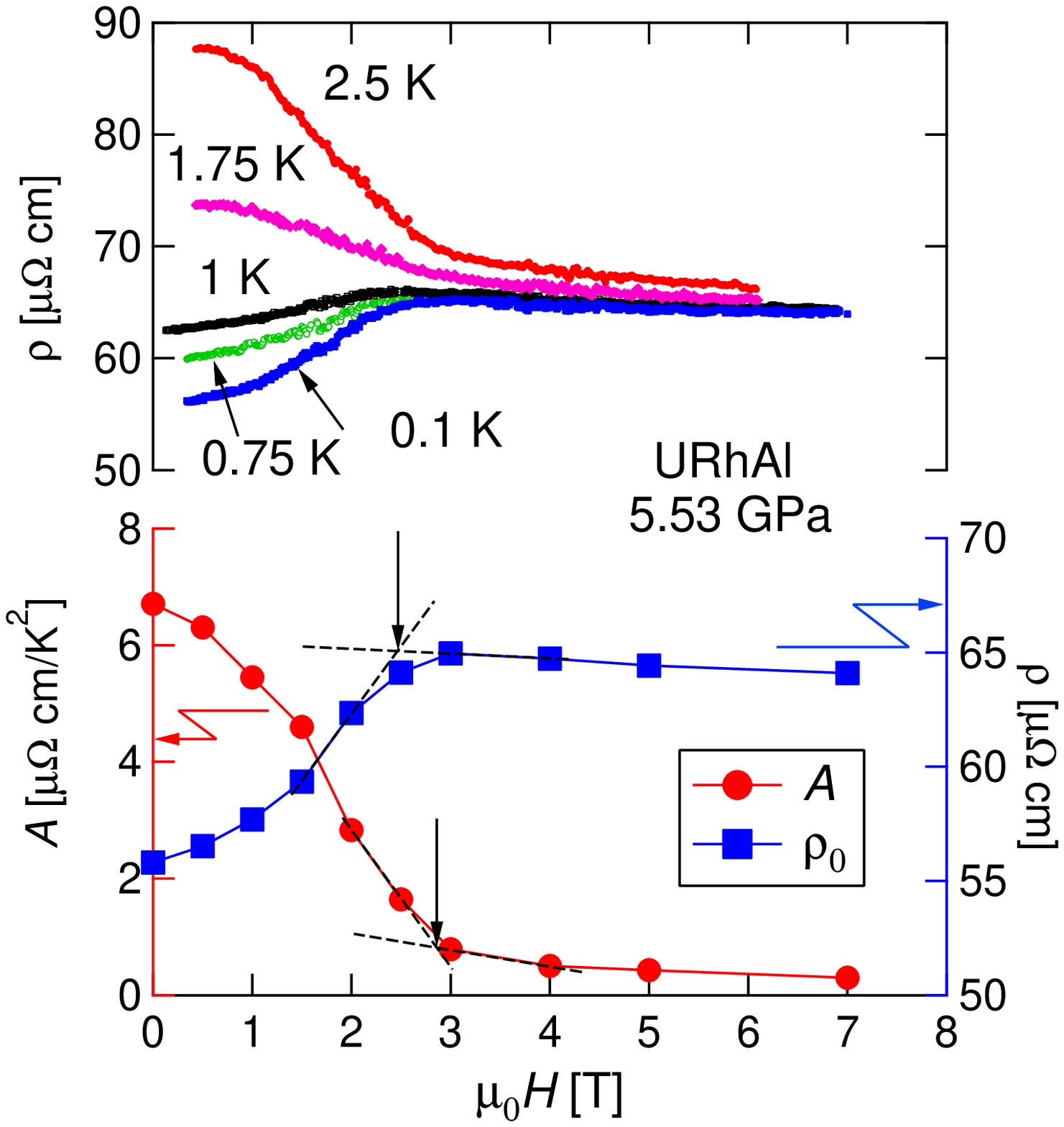}
\caption{(Color online)　Magnetic field dependence of resistivity of URhAl for the sample $\#1$, measured at 5.53 GPa. 
 The dashed lines are guides to the eyes.
 \color{black}
}
\end{figure}

The experimentally observed large enhancement of the $A$-coefficient 
 suggests a large mass enhancement due to spin-fluctuation effects
 and/or a variation of  Fermi surface.
Generally, large fluctuation effects occur  
 for a 2nd-order phase transition with a divergence of the correlation length of the magnetic order.
In contrast, such an effect would not be expected for a 1st-order phase transition.
Nevertheless, if the transition near the $P_{\mathrm{c} }$ is only weakly 1st-order and the drop of the FM moment at the 
 FM-PM phase transition is very small, the critical behavior becomes similar to that of a  QCP, 
 and 
 then a large maximum in the $A$-coefficient may emerge due to the increase of 
 correlation length as $T \rightarrow 0$.

Figure 5 shows the $A$-coefficient and the residual resistivity as a function of magnetic field.
At 4.5 GPa, $A(H)$ value is very small, and $A(H)$ monotonically decreases with increasing field.
At  5.0 GPa, the $A$-coefficient begins to increase in zero and low fields, 
 and $A(H)$ is suddenly suppressed by magnetic field of $\sim$1 T.
At around 5.2-5.5 GPa, the $A$-coefficient is very large in zero field and 
 remains large up to  1-1.5 T, then rapidly decreases at high fields (1.5-2 T).
At  6.0 GPa, the decrease of $A(H)$ occurs at higher field near 3 T.
At  6.9 and 7.3 GPa, 
 the value of $A$-coefficient at 0 T becomes about half of the $A$ value at 5.5 GPa,
 and after  showing a slight maximum at around 2 T,  it monotonically decreases with increasing field.
At  6.9 and  7.3 GPa, $\rho_{0}(H)$ increases with increasing field, 
and shows a smooth maximum at around 3 T.

To search  for the FM wing structure,
 we look at the  magnetic field dependence of the resistivity [$\rho(H)$] under high pressure.
Figure 6 shows $\rho(H)$ under 5.53 GPa at 2.5, 1.75, 1, 0.75, and 0.1 K 
 with   $A(H)$ and  $\rho_{0}(H)$ obtained from the temperature dependence of $\rho(T) = \rho_{0} + AT^2$.
$\rho(H)$ curve bends at around  2.5-3 T for each temperature.
We define the anomaly at $H_{\mathrm{m} }$  at $T \rightarrow 0$ from  $A(H)$ and  $\rho_{0}(H)$  as indicated by the  arrows in  Fig. 6.
At  the low-field region below $H_{\mathrm{m}}$,  
 the $A$-coefficient is very large compared to the high-field region above $H_{\mathrm{m} }$.
On the other hand,
 the high-field region above  $H_{\mathrm{m} }$
  corresponds to the FM side, 
 where the resistivity obeys $\rho(T) = \rho_{0} + AT^2$ with the small $A$-coefficient.

\begin{figure}[!htb] %
\includegraphics[width=8.6cm]{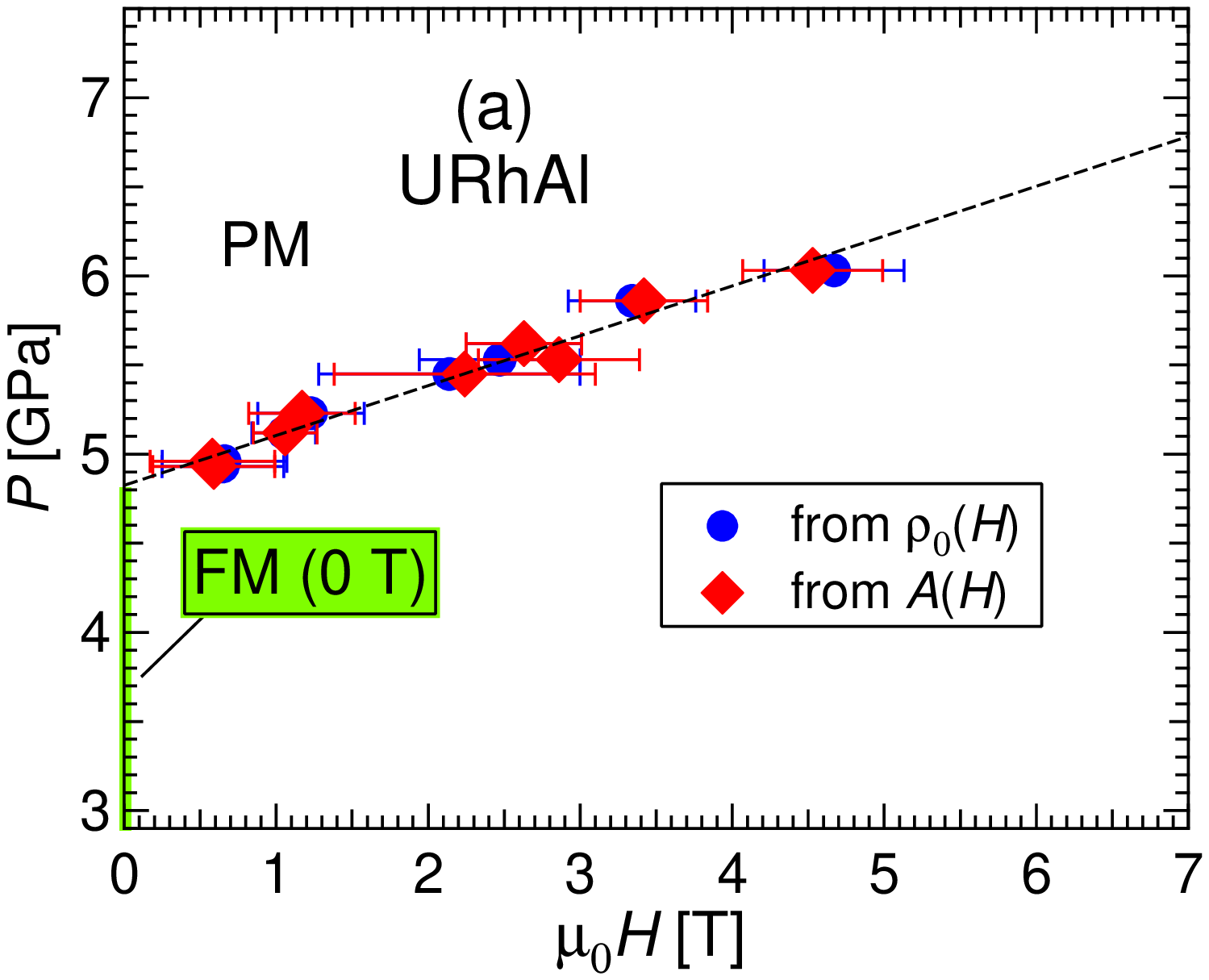}
\includegraphics[width=8.6cm]{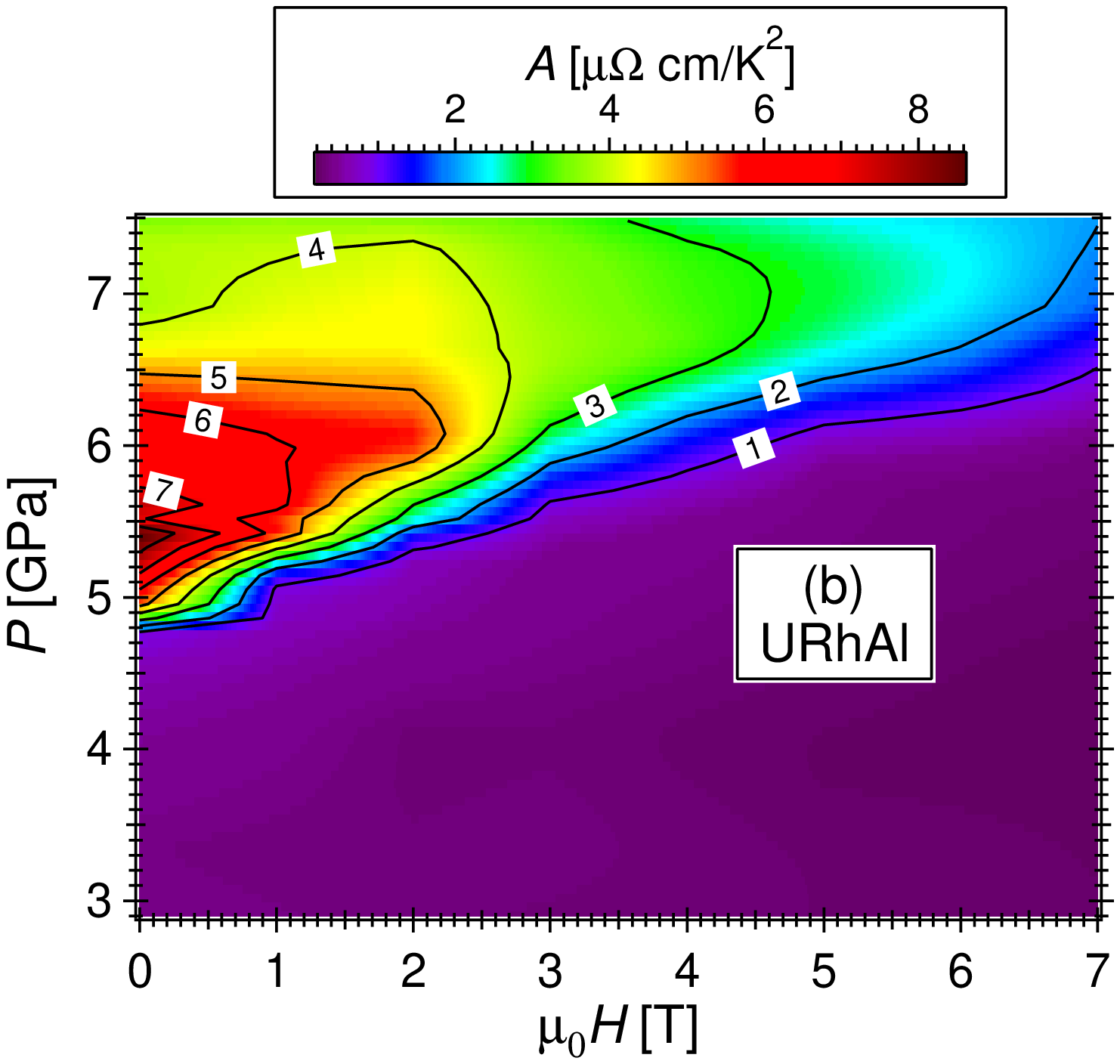}
\caption{(Color online)  (a) Plot of the observed anomalies  in $A(H)$ and $\rho(H)$ of URhAl on the  
$P$-$H$ phase diagram   for the sample $\#1$. 
   The dashed line indicates the result of liner fitting.
 \color{black} 
 (b)
Contour plot of the $A$-coefficient  of resistivity on URhAl on the $P$-$H$ phase diagram, obtained from the expression  $\rho(T) = \rho_0 + AT^2$
  for the sample $\#1$. 
}
\end{figure}

The anomaly in $\rho(H)$ curves  supports the presence of  FM wing structure in URhAl.  
In Fig. 7(a), we plot the  $P$-$H$ phase diagram for  $H_{\mathrm{m} }$
 obtained from the magnetic field dependences of $A$-coefficient and $\rho_{0}$. 
We obtain the relation $\mu_{0} dH_{\mathrm{m}}/dP \sim$ 3.5 $\pm$0.1 T/GPa.
Then 
 we estimate the TCP at $P_{\mathrm{TCP} } \sim$ 4.8-4.9 GPa, when $H_{\mathrm{m}} \rightarrow 0$, 
 using the value of $\mu_{0} dH_{\mathrm{m}}/dP $.
 In 
 UCoAl,  which has the same crystal structure as URhAl,
  a clear 1st-order metamagnetic transition  is seen at $H_{\mathrm{m}}$ due to  
 the FM wing. 
According to high pressure studies 
 \cite{Aoki_JPSJ_2011}, 
 the metamagnetic transition field of UCoAl varies as 
 $\mu_{0} dH_{\mathrm{m}}/dP \sim$ 3.5 T/GPa
 \cite{Aoki_JPSJ_2011},
 which  is very similar to that  of URhAl.
We summarize the values of $P_{\mathrm{TCP}} $ and   $\mu_{0} dH_{\mathrm{m}}/dP $ in Table. II.

\begin{figure}[!htb] %
\includegraphics[width=8cm]{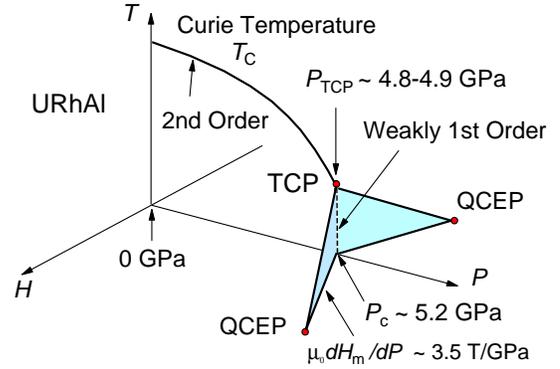}
\caption{(Color online)  Schematic $T$-$P$-$H$ phase diagram of the FM wings in URhAl [See also the top panel of Fig. 2 and Fig. 7(a)].
}
\end{figure}

\begin{table}[!htb]
\caption{The values of  $P_{\mathrm{TCP}}$ and 
 the slope of the FM wing, i.e. $\mu_{0} dH_{\mathrm{m}}/dP$ 
 for URhAl and UCoAl \cite{Aoki_JPSJ_2011}. 
 }
\begin{ruledtabular}
\begin{tabular}{lcr}
        & $P_{\mathrm{TCP}} $ [GPa]  &   $\mu_{0} dH_{\mathrm{m} }/dP$  [T/GPa] \\
\hline
URhAl   &   4.8-4.9     &  3.5 $\pm$0.1  \\
UCoAl   &  $-$0.2      &  3.5   \\
\end{tabular}
\end{ruledtabular}
\end{table}

When we cross the FM wing, we expect the 1st-order PM-FM phase 
 transition  at  $H_{\mathrm{m} }$.
 At the 1st-order transition in UCoAl,  
 the $A$-coefficient shows a step-like behavior as a function of magnetic field
 \cite{Aoki_JPSJ_2011}.
On the other hand, 
 the  step-like behavior in the $A$-coefficient  of URhAl 
  near  $H_{\mathrm{m}}$   is rather  broad.
This may indicate that  the transition at $H_{\mathrm{m} }$ is  weakly  1st-order  in URhAl.
However, the sample quality can be the origin  of the broadness of the transition.


We shall compare  $A(H)$ for URhAl with that for UCoAl.
For UCoAl, a step-like behavior in $A(H)$ at  $H_{\mathrm{m}}$
 is seen 
 under low pressure region below 0.54 GPa.
For 0.54 GPa, 
 the difference of pressure from the TCP ($\sim -$0.2 GPa ) is 
 estimated to be  $\delta P \equiv P - P_{\mathrm{TCP} } \sim $ 0.74 GPa.
Since we estimate $P_{\mathrm{TCP} } \sim$ 4.8 GPa for URhAl in the present work,
 the pressure of 0.54 GPa in UCoAl 
 may correspond to a pressure of 
 $4.8 + \delta P  \sim$ 5.54 GPa in URhAl.
At $\sim$ 5.5 GPa, we obtain $H_{\mathrm{m}} \sim 2.5$ T for URhAl from Fig. 7(a),
 which is close to the value of $H_{\mathrm{m} }$ for UCoAl at 0.54 GPa.

In contrast, 
 the enhancement of the $A$-coefficient in URhAl
 is much larger than the value in UCoAl (see, Table I),
 suggesting that the density of states (mass enhancement and/or the change of Fermi surface)  
 near the QPT is more drastic in URhAl than UCoAl. 
In URhAl, the $A$-coefficient  below  $H_{\mathrm{m} }$ at $\sim$ 5.0 GPa 
  is about 
20 times larger than the $A$-coefficient above $H_{\mathrm{m} }$ (Fig. 5).
 In UCoAl, on the other hand,  
 the $A$-coefficient in the PM state below  $H_{\mathrm{m} }$
  is about only 2 times larger than 
  the $A$-coefficient above  $H_{\mathrm{m} }$ at 0 and 0.54 GPa 
 \cite{Aoki_JPSJ_2011}.

 The difference of the $A$-coefficient between URhAl and UCoAl may be related to that of the magnetic ordered moments; 
  the FM ordered  moment in URhAl
  is 3 times larger ($\sim$ 0.9 $\mu_{\mathrm{B} }$/U ) 
 than the magnetic-field-induced FM moment ($\sim$ 0.3 $\mu_{\mathrm{B} }$/U) in UCoAl. 
As shown theoretically by Yamada \cite{Yamada_PRB_1993},
  thermally fluctuating magnetic moments enhance the scattering of quasiparticles in the PM state,
 and may cause such a large $A$-coefficient. 
At present, the ordered moment near $P_{\mathrm{c}}$ has not yet been studied for URhAl, 
 so further studies are necessary to clarify this point.

The behavior of the $A$-coefficient changes 
  in association with the wing structure. 
To see the relationship between the enhancement of $A$-coefficient and the FM wing,
 it is intriguing to see  the contour plot of $A$-coefficient on the $P$-$H$ phase diagram.
Figure 7(b) shows the contour plot of the $A$-coefficient  on the $P$-$H$ 
 phase diagram, obtained from   $\rho(T) = \rho_{0} + AT^2$.
The red-colored  region in this  plot shows the enhancement of the $A$-coefficient,
 whereas the purple-colored region shows the small  $A$-coefficient.
The $A$-coefficient is  largest at around  5.2-5.5 GPa in zero field.
With increasing pressure and magnetic field,
 the $A$-coefficient is suppressed.
The large enhancement of $A$-coefficient occurs outside of the FM wing  (red region).

In Fig. 8, we plot the schematic $T$-$P$-$H$ phase diagram of the FM wings in URhAl
 [See also the top panel of Fig. 2 and Fig. 7(a)].
The theoretically suggested 1st-order wing planes  terminate at the QCEP 
 at  zero temperature  in a finite field.
In UCoAl 
    the magnetic-field dependence of $A(H)$ coefficient shows a sharp maximum  at the PM-FM transition 
 near the QCEP  ($P \sim$ 1.5 GPa in $H \sim$7 T)
 \cite{Aoki_JPSJ_2011, TristanCombier_Dr_Thesis_2014}.
Such a field enhancement 
 in  $A(H)$ was not observed in the present work,
 suggesting that the 
 QCEP of URhAl may  exist above $\sim$7 T.
Alternatively, the interplay between 
 spin fluctuation and Fermi-surface instability can 
 lead to complex phenomena as discussed later.

\begin{figure}[!htb]
\includegraphics[width=8.6cm]{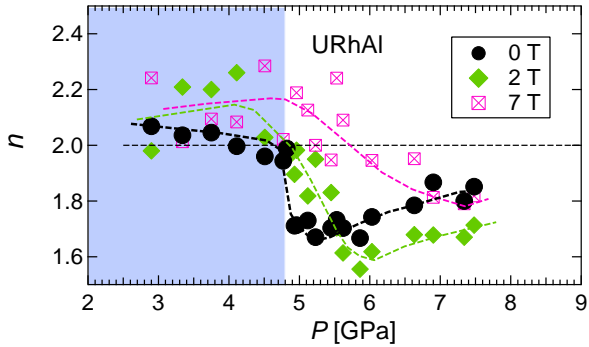}
\caption{ (Color online) Pressure dependence   
  of the exponent($n$) of resistivity [$\rho(T) = \rho'_0 + A'T^n$] for  URhAl (sample $\#1$) at 0,  2, and 7 T. 
The dashed lines are guides to the eyes.
}
\end{figure}

As close to the critical pressure, the temperature range for Fermi-liquid regime [$\rho(T) = \rho_0 + AT^2$] is found 
 to be  very small ($T^{*} \sim $ 0.4 K), 
 an alternative description is 
  to focus on the 
NFL behavior. 
 We analyzed the resistivity data by the expression, $\rho(T) = \rho'_0 + A'T^n$.
 Here, the maximum temperature for the NFL regime $\rho(T) = \rho'_0 + A'T^n$, 
 i.e., $T^{**}  \sim 2.2$ K at $\sim$ $P_{\mathrm{c} }$. 
\color{black}
 Figure 9 \color{black} shows 
  the 
 pressure dependence 
 of the exponent of resistivity ($n$).
As seen in  
 Fig. 9\color{black}, at 0 T 
 the exponent $n$ is about 2 below 4.8 GPa,
 whereas $n(P)$ decreases with a step-like variation to a minimum at around  $\sim $5.0-6.0 GPa.
At around $\sim$5.0-6.0 GPa, the values of $n$ are about 1.6-1.8.
At  2 T and 7 T,  
 the step-like behavior and the minimum of the exponent $n(P)$ shift  to higher pressure regions.
At 7 T, the dip of $n(P)$ at around $\sim$ 7.0 GPa becomes shallow and broad, 
 and  the value of $n(P)$ becomes slightly larger than  at 0 T and 2 T.  

The  exponent $n$ (Fig. 9)
 \color{black}
 varies as a function of pressure  corresponding to 
 the behavior of the $A$-coefficient (Fig. 4).
Above  $P_{\mathrm{c} }$,  the exponent $n$ is nearly 1.6-1.7, which is close to the value ($n = $ 5/3)
  suggested from the SCR theory for three-dimensional FM  
 spin fluctuation near a QCP.
However,  this NFL behavior seems to conflict with the presence of the FM 1st-order wing structure 
  in URhAl.
A  weakly 1st-order nature at the FM wing may 
 explain the NFL behavior in the resistivity.

Hereafter, we note that 
 the critical behavior around $P_{\mathrm{c}}$ in URhAl is 
 different  from the theoretical suggestion  for a 2nd-order FM  QCP.
It is suggested that 
 the ratio of  $T^{*}$ for the Fermi-liquid regime to  
 $T^{**}$ for the NFL regime
  is enhanced as  $T^{**}/T^{*} \propto (P - P_{\mathrm{c} })^{-3/4}$
 as approaching $P_{\mathrm{c}}$ for a 2nd-order FM QCP 
 \cite{Millis_PRB_1993, Flouquet_ArXiv_2005}.
For URhAl,   $T^{**}$ does not change clearly, i.e.,  $ T^{**} \sim 2.1 \pm 0.2$ K.
In addition, 
$T^{*}(P)$ is almost linear (Fig. 2). 
These experimental 
 results for URhAl suggest  
 that the spin-fluctuation 
 effects cannot be explained simply by 
   the 2nd-order FM QCP again.


In URhAl, the NFL properties (see  Fig, 9)  are 
 observed far above  $P_{\mathrm{c}}$,
  and the pressure domain of the enhancement of the  $A$-coefficient
 appears quite asymmetric around  $P_{\mathrm{c} }$.
Furthermore the enhancement of the  $A$-coefficient extends over a large $P$ window (5.5-7 GPa).
Then the key question is if 
 the switch from the FM state to the PM state simply occurs 
 at  $P_{\mathrm{c}}$  or if there is a mark of a new pressure-induced phase intermediate between the FM and the PM phases.

Recently 
 it has been shown theoretically that 
  another   new phase possibly  stabilizes  
  near  the  FM-PM  QPT
 \cite{ Maslov_PRB_2009, Chubukov_PRL_2009, Conduit_PRL_2009, 
Karahasanovic_PRB_2012, Thomson_PRB_2013,  Pedder_PRB_2013}; 
 if there are  quantum fluctuations  in terms of fermionic 
 particle-hole excitations  on the Fermi surface, 
 some deformations of the Fermi surface  enhance the phase space available for 
   the quantum fluctuations,  and
 such a Fermi-surface instability  causes 
another type of  ordering to   
 gain the free energy of the system
 \cite{Conduit_PRL_2009,  
 Karahasanovic_PRB_2012, Thomson_PRB_2013,  Pedder_PRB_2013}.

It has been shown that 
 two cases of new ordering are possible near the FM-PM QPT:
 (i) spiral  magnetic phase, and (ii) spin-nematic phase 
\cite{Karahasanovic_PRB_2012}.
The energy scale of the spin-nematic phase transition  is 
 almost 10 times smaller than the case of spiral  magnetic phase \cite{Karahasanovic_PRB_2012},
 therefore a spiral  magnetic phase
  might be more likely to occur.
A spiral magnetic phase emerges \textit{below} the  TCP 
 as an \textit{ intermediate state}  
 between   the uniform FM and the PM  states \cite{Karahasanovic_PRB_2012, Thomson_PRB_2013}.
 The  transition between  the uniform FM  and the spiral magnetic states
 occurs as a Lifshitz transition,
 whereas the  transition between   
 the spiral magnetic state and the PM state 
 is of  1st-order 
 \cite{Karahasanovic_PRB_2012}.
Interestingly,  an anisotropic dispersion for electron band changes 
  the nature of the spiral-to-PM phase transition, 
  and the transition possibly becomes   2nd-order   \cite{Karahasanovic_PRB_2012}.

The possible presence of 
 the intermediate new phase might explain why 
 we cannot see a clear 1st-order transition between the FM and the PM  states 
  above  $P_{\mathrm{c} }$, 
 different from the case of UCoAl.
 In order to explore such a new phase around  the QPT,
 further experimental studies are required.
In particular, 
 measurements of thermodynamic quantities and 
 observation of Fermi-surface change through  $P_{\mathrm{c} }$  for URhAl, 
 though experimentally challenging, 
 would  deepen the understanding 
 the Fermi-surface instability and the nature of the FM QPT.

In URhAl, no superconductivity was observed 
 under high-pressure up to 7.5 GPa at low temperature down to $\sim$ 0.1 K.
At present we cannot rule out 
 the possibility that 
 the sample quality  affects the emergence of superconductivity and the superconducting transition temperature 
 is too low to be detected.
However, even if a superconductivity does not occur,
 the FM system would resolve the instability due to the quantum fluctuation at $T$ $\rightarrow$ 0
 by  the occurrence of another new phase associated with the Fermi-surface instability 
 as mentioned just above.

It is interesting to consider why the intermediate phase possibly appears in URhAl. 
One may consider that
 the lack of local inversion center and/or the quasi-Kagome lattice in ZrNiAl-type structure can induce the intermediate phase. 
However,  
 the ZrNiAl-type hexagonal symmetry structure ($P\bar{6}2m$: $D_{3h}^{3}$) does not lead 
 to Dzyaloshinskii-Moriya interaction
 \cite{Kataoka_JPSJ_1981}, which could induce a helimagnetic order \cite{Dzyaloshinskii_JETP_1964}. 
Such an intermediate phase has not yet been observed in UCoAl, which has the same ZrNiAl-type  structure, around the PM-FM phase transition induced by uniaxial stress along the $c$-axis 
 \cite{Ishii_PhysicaB_2003, Karube_JPSJ_2014, YShimizu_JPSJ_2015}.
 The relationship between the crystal structure and the occurrence of the intermediate phase remains open question. The authors in Ref.  \cite{Karahasanovic_PRB_2012}   
  suggest that the intermediate phase may generally occur even for a simple spherical Fermi surface due to the Fermi-surface instability accompanying  the quantum fluctuations (particle-hole excitations on the Fermi surface). 
 As seen in Fig. 9, 
 the NFL behavior of the resistivity is remarkable in URhAl  far above $P_{\mathrm{c}}$. 
 Such strong quantum-fluctuation effects near the FM-PM QPT may invoke the intermediate phase in this material.
\color{black}

\color{black}
\section{Conclusion}
 
The quantum criticality of 
 the  three-dimensional-Ising-type itinerant FM compound URhAl was studied 
 by low-temperature resistivity measurements under high pressure up to  7.5 GPa. 
The Curie temperature is suppressed with increasing pressure, and  suddenly disappears above 5.0 GPa.
Our resistivity results suggest the FM critical pressure of $\sim$ 5.2 GPa. 
Above 5.2 GPa, the ground state is not FM, and  
  the $A$-coefficient is largely enhanced  
  at  around 5.2-5.5 GPa  in zero and low-field region.
The characteristics of the temperature  and the magnetic-field dependences of the
 resistivity  may be  consistent with the presence of a FM wing structure 
 with an estimated TCP at 4.8-4.9 GPa.
At least with the present quality of the crystal the 1st-order phase transition appears weak.
The resistivity  shows the NFL behavior above  5.0 GPa up to  7.5 GPa.
URhAl  may be a  material in  which
 the switch from the FM state to the PM state occurs through an intermediate phase
 around  the  QPT.

\section*{ACKNOWLEDGMENT}

We would like to thank S. Kambe, G. Knebel,  
 K. Ishida, Y. Tada,  K. Hattori, 
 S. Hoshino, and Y. Ikeda for valuable
 discussions and helpful comments.
This work was supported by ERC starting grant New Heavy Fermion,
 KAKENHI, REIMEI, ICC-IMR, and ANR project PRINCESS.

\bibliography{apssamp}


\end{document}